# Data-driven femtosecond optical soliton excitations and parameters discovery of the high-order NLSE using the PINN

Yin Fang [#], Gang-Zhou Wu [#], Yue-Yue Wang [*] and Chao-Qing Dai [*]

*College of Sciences, Zhejiang A&F University, Lin'an, Zhejiang 311300, P. R. China*

**Abstract.** We use the physics-informed neural network to solve a variety of femtosecond optical soliton solutions of the high order nonlinear Schrödinger equation, including one-soliton solution, two-soliton solution, rogue wave solution, W-soliton solution and M-soliton solution. The prediction error for one-soliton, W-soliton and M-soliton is smaller. As the prediction distance increases, the prediction error will gradually increase. The unknown physical parameters of the high order nonlinear Schrödinger equation are studied by using rogue wave solutions as data sets. The neural network is optimized from three aspects including the number of layers of the neural network, the number of neurons, and the sampling points. Compared with previous research, our error is greatly reduced. This is not a replacement for the traditional numerical method, but hopefully to open up new ideas.

## 1. Introduction

With the development of plasma physics, optical fiber communication and other disciplines, the nonlinear problem has attracted more and more attention [1-3]. These nonlinear phenomena can be described by nonlinear partial differential equations(NPDEs). Solving the NPDEs, one can reveal the nature of nonlinear phenomena [4,5]. The nonlinear Schrodinger equation(NLSE) can be used to describe the propagation of optical solitons in optical fibers. In recent years, the research of ultrashort pulse lasers has become a popular direction. But for describing the femtosecond light pulses propagating in optical fibers, the standard NLSE becomes insufficient. High-order effects, such as third-order dispersion(TOD) and nonlinear response effects, will play a crucial part in the propagation of ultrashort pulses, such as femtosecond pulses. To understand this phenomenon, the high order nonlinear Schrödinger equation (HNLSE) was proposed [6]. Recently, neural networks have also been used to study NLSE and other partial differential equations [7-9], however, they have not been extended to study optical soliton excitations of HNLSE.

In recent decades, the rapid development of data processing and computing capabilities has promoted the application of deep learning in the field of data mining, such as face recognition, machine translation, biomedical analysis, traffic prediction, and autonomous driving [10-12].

However, the acquisition of data is a huge challenge. How to obtain information effectively and accurately when part of the data is missing is an urgent and arduous task [13,14]. At the same time, due to the lack of sample data and poor robustness, the results obtained by the traditional maximum likelihood method are often unreliable. In fact, it seems impossible to use the maximum likelihood method of input and output data to draw conclusions related to the laws of physics, especially for high-dimensional problems [13,15]. Researchers speculate that defects in the prior laws related to physical systems may be one of the main reasons. Therefore, many researchers try to use the maximum likelihood algorithm and some physical laws to improve the accuracy of the unknown solution of the physical model [16,17].

Recently, Raissi et al. fully integrated the information related to the physical system into the neural network, using the maximum likelihood technique to establish a deep learning method for physical constraints [18,19], called the physics-informed neural network (PINN) and its related improvements [20,21]. It is a method that is not only suitable for solving the forward problem of NPDEs, but also suitable for solving the inverse problem of NPDEs [22,23]. The PINN neural network can obtain very accurate solutions with less data, and has good robustness [24]. At the same time, the physical information is expressed by differential equations, which also provides good physical meaning for predicting the solution [25]. This paper proposes a data-driven algorithm with high computational efficiency to derive the solution of more complex NPDEs.

The general form of HNLSE is as follows[26]

$$iQ_t + \lambda_1 Q_{xx} + \lambda_2 |Q|^2 Q + i[\lambda_3 Q_{xxx} + \lambda_4 (|Q|^2 Q)_x + \lambda_5 Q(|Q|^2)_x] = 0, \qquad (1)$$

where $Q$ is a complex function related to the delay time $x$ and longitudinal propagation distance $t$ in the Eq. (1). $\lambda_1, \lambda_2, \lambda_3, \lambda_4$ and $\lambda_5$ are the real parameters respectively related to group velocity dispersion, Kerr nonlinearity, TOD, self-steepening, and self-frequency coming from stimulated Raman scattering. If a picosecond optical pulse is studied, then $\lambda_3, \lambda_4$ and $\lambda_5$ in Eq. (1) are all zero, and the HNLSE degenerates to the standard NLSE. If the duration of the pulse is less than 100fs, $\lambda_3$, $\lambda_4$ and $\lambda_5$ will not be zero.

The main novelty of this paper is as follows. (i) The HNLSE is firstly studied by the PINN; (ii) Five femtosecond optical soliton excitations including one-soliton, two-soliton, rogue wave and W-soliton, M-soliton solutions of the HNLSE are trained by the PINN. Compared with the previous research[24], the error in this paper is smaller. (iii) In the inverse problem, the structure of the neural network is optimized from three aspects: the number of layers of the neural network, the number of neurons, and the sampling points. These research perspectives are not considered by previous literatures.

## 2. PINN method

Neural networks have the properties of general function approximators and can approximate any function. Therefore, it can be directly used to deal with nonlinear problems, avoiding limitations such as preset, linearization, or local time stepping. In this paper, the improved PINN method is used to reconstruct the dynamic characteristics of the HNLSE, and the parameters of the

HNLSE are obtained by data driving. The general form of (1+1)-dimensional complex NPDE is as follows

$$Q_t + N(\lambda, Q, Q_x, Q_{xxx}, \cdots,) = 0, \quad x \in (x_1, x_2), \quad t \in (t_1, t_2), \tag{2}$$

where $N$ is a combination of linear and nonlinear terms about $Q$. Eq. (1) is the underlying physical constraint, thus forming one multilayer feed forward neural network $\tilde{Q}(t,x)$ and PINN $f(t,x)$ that share parameters with each other (such as, scaling factors, weights, and deviations). Neural networks learn shared parameters by minimizing the mean square error (MSE) caused by the initial boundary value conditions associated with the feed forward neural network $\tilde{Q}(t,x)$ and PINN $f(t,x)$. Since $Q(t,x)$ is a complex valued number, we need to separate the real and imaginary parts of $Q(t,x)$, whose real part is $g(t,x)$ and imaginary part is $h(t,x)$. From Eq. (2), we get

$$f_g := g_t + N_g(\lambda, g, g_x, g_{xxx}, \cdots,), \tag{3}$$

$$f_h := h_t + N_h(\lambda, h, h_x, h_{xxx}, \cdots,). \tag{4}$$

The Predicted solution $\hat{Q}(t,x)$ ($\hat{Q} = \sqrt{g^2 + h^2}$) is embedded in PINN through Eqs. (3) and (4), which are used as physical constraints to prevent the over-fitting phenomenon of the neural network, as well as to predict the solution, which provides a good physical interpretation. The loss function $\Gamma$ has the following form

$$\Gamma = \Gamma_g + \Gamma_h + \Gamma_{f_g} + \Gamma_{f_h} \tag{5}$$

in which

$$\Gamma_g = \frac{1}{N_q} \sum_{l=1}^{N_q} \left| g(t_g^l, x_g^l) - g^l \right|^2, \tag{6}$$

$$\Gamma_h = \frac{1}{N_q} \sum_{l=1}^{N_q} \left| h(t_h^l, x_h^l) - h^l \right|^2, \tag{7}$$

$$\Gamma_{f_g} = \frac{1}{N_f} \sum_{j=1}^{N_f} \left| f_g(t_{f_g}^j, x_{f_g}^j) \right|^2, \tag{8}$$

$$\Gamma_{f_h} = \frac{1}{N_f} \sum_{j=1}^{N_f} \left| f_h(t_{f_h}^j, x_{f_h}^j) \right|^2. \tag{9}$$

Here, the initial value and boundary value data about $Q(t,x)$ are obtained from $\{t_g^l, x_g^l, g^l\}_{l=1}^{N_q}$, and $\{t_h^l, x_h^l, h^l\}_{l=1}^{N_q}$, In the same way, the collocation points of $f_h(t,x)$ and $f_g(t,x)$ are specified by $\{t_{f_g}^j, x_{f_g}^j\}_{j=1}^{N_f}$ and $\{t_{f_h}^j, x_{f_h}^j\}_{j=1}^{N_f}$. In this paper, $N_q = 100$, $N_f = 10000$. In PINN, we choose to use the Adam optimizer to optimize the loss functions Eq.(5), combining the advantages of AdaGrad and RMSProp, which has become one of the mainstream optimizers. In addition, we choose the hyperbolic tangent function tanh as the activation function [27].

## 3. Data-driven optical soliton solutions

In this section, we investigate HNLSE by using a PINN, and train five femtosecond optical soliton excitations including one-soliton, two-soliton, rogue wave and W-soliton, M-soliton solutions via PINN.

### 3.1 One-soliton solution

We take the parameter $\lambda_1 = 0.5, \lambda_2 = 1, \lambda_3 = -0.18, \lambda_4 = -1.08$ and $\lambda_5 = 1.08$ in Eq. (1), and the exact solution of the one soliton is as follows [28]

$$Q(t,x) = -0.5\operatorname{sech}(0.5x + 0.0225t)e^{0.125it}, \quad x \in [-15,15], \quad t \in [0,5]. \tag{10}$$

To use the PINN deep learning, we let the initial condition

$$Q(0,x) = -0.5\operatorname{sech}(0.5x), \quad x \in [-15,15], \tag{11}$$

and the Dirichlet-Neumann periodic boundary condition

$$Q(t,-15) = Q(t,15), \quad t \in [0,5]. \tag{12}$$

The sampling points set is obtained by means of pseudo-spectral method with space-time region $(x,t) \in [-15,15] \times [0,5]$, and the exact one-soliton solution is discretized into $[256 \times 201]$ data points. The initial and Dirichlet periodic boundary data are obtained by Latin hypercube sampling [29]. The sampling points set used in the 9-layer neural networks consists of points $N_q = 100$ randomly sampled from the initial data given by Eq. (11), and the periodic boundary data given by Eq. (12), and collocation points $N_f = 10000$ for the PINN $f(t,x)$ given by Eq. (1). In addition, the MSE loss function given by Eq. (5) is learned by using a PINN with 7 hidden layers of 30 neurons in each layer. After 4000 iterations of learning, the network achieved a relative error $L_2$ as $L_2 = [\|\hat{Q}(t,x) - Q(t,x)\|]/Q(t,x)$ of $9.125253 \times 10^{-3}$ in about $5545.7248$ seconds.

Fig. 1 exhibits reconstructed one-soliton space-time dynamics. In Fig. (a), the sampling point configuration $|Q(t,x)| = \sqrt{g^2(t,x) + h^2(t,x)}$ of the initial and boundary is clearly analyzed. Fig. 1(b) reconstructs the space-time dynamics of one-soliton for the HNLSE. We compare the predicted solution with exact solution for three different distances $t = 1.5, 2.5, 4$ in Fig. 1 (c). Fig. 1(d) shows an Error $Er = |\hat{Q}(t,x) - Q(t,x)|$ between the predicted solution $\hat{Q}(t,x)$ and exact solution $Q(t,x)$.

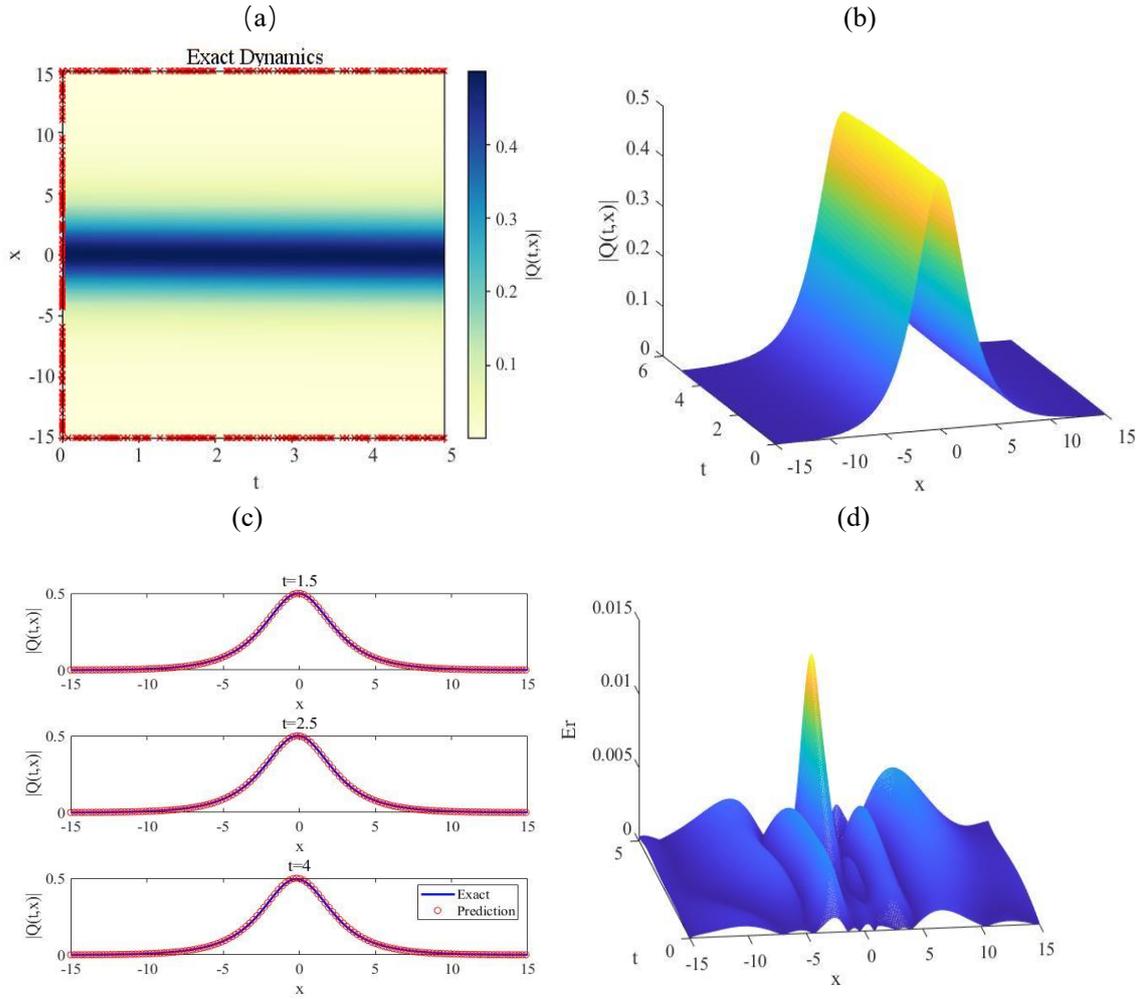

**Figure 1.** The one-soliton solution $Q(t,x)$: (a) Exact one-soliton solution $|Q(t,x)|$ with the boundary and initial sampling points depicted by the cross symbol; (b) Reconstructed one-soliton space-time dynamics; (c) Comparison between exact and predicted solutions at three distances, with the red hollow dots as the predicted values and the blue solid lines as the exact values; (d) Three-dimensional stereogram of error.

### 3.2 two-soliton solution

Here we consider two soliton interaction and take the parameter $\lambda_1 = 0.5$, $\lambda_2 = 1$, $\lambda_3 = \frac{1}{30}$, $\lambda_4 = 0.2$ and $\lambda_5 = -0.2$ in Eq. (1). Exact solution of the two solitons interaction is given in Ref.[30] as follows

$$Q(t,x) = \frac{0.02205\cosh(1.1x + 0.1331t)e^{i0.5t} - 0.024255\cosh(x + 0.1t)e^{i0.605t}}{0.0025\cosh(2.1x + 0.2331t) + 1.1025\cosh(0.1x + 0.0331t) - 1.1\cos(0.105t)}. \tag{13}$$

Among the region $x \in [-10,10]$, $t \in [-5,5]$, we get the initial condition

$$Q(-5,x) = \frac{0.02205\cosh(1.1x - 0.6655)e^{-2.5i} - 0.024255\cosh(x + 0.5)e^{-3.025i}}{0.0025\cosh(2.1x - 1.1655) + 1.1025\cosh(0.1x - 0.1655) - 1.1\cos(-0.525)}, \tag{14}$$

and the Dirichlet-Neumann periodic boundary condition

$$Q(t,-10) = Q(t,10), \quad t \in [-5,5]. \tag{15}$$

The sampling points set is obtained by means of pseudo-spectral method with space-time region $(x,t) \in [-10,10] \times [-5,5]$, and the exact two-soliton solution is discretized into $[512 \times 401]$ data points. Similar to the procedure of one-soliton, after $4000$ iterations of learning, the network achieved a relative error $L_2$ of $1.633679 \times 10^{-2}$ in about $11801.0978$ seconds.

Fig. 2 exhibits reconstructed two-soliton space-time dynamics. In Fig. 2(a), the sampling point configuration of the initial and boundary is clearly analyzed. Fig. 2(b) reconstructs the space-time dynamics of the two solitons of the HNLSE. In Fig. 2 (c), we compare the predicted solution with exact solution at three different distances $t = -2.5, 0, 2.5$. Fig. 2(d) shows the Error $Er = |\hat{Q}(t,x) - Q(t,x)|$ between the predicted and exact solutions. It can be seen from the density map that the optimized PINN prediction results have good accuracy in the whole time and space domains.

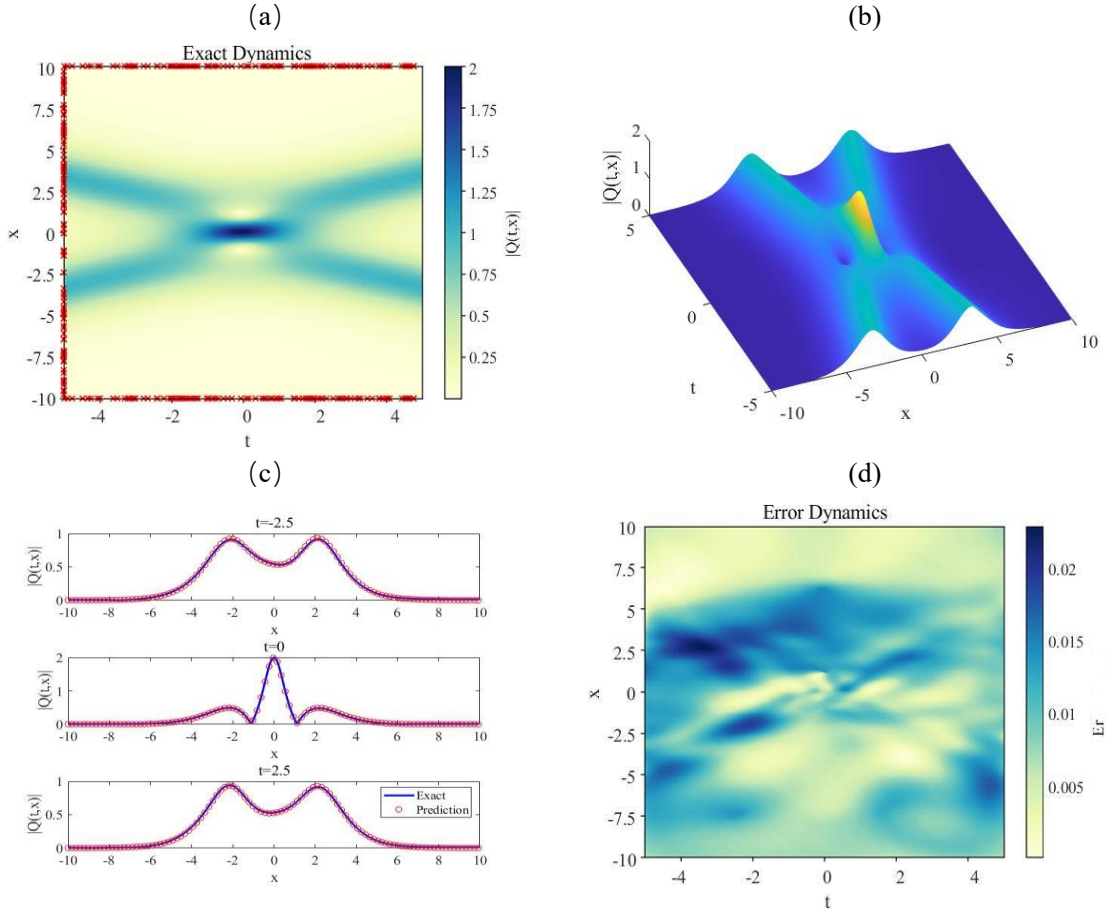

**Figure 2.** Two-soliton solution $Q(t,x)$: (a) Exact two-soliton solution $|Q(t,x)|$ with the boundary and initial sampling points; (b) Reconstructed two-soliton space-time dynamics; (c) Comparison between exact and predicted solutions at three distances, with the red hollow dots as the predicted values and the blue solid lines as the exact values; (d) The error density plot.

**3.3 Rogue wave solution**

We take the parameter $\lambda_1 = 0.5$, $\lambda_2 = 1$, $\lambda_3 = 0.1$, $\lambda_4 = 0.6$ and $\lambda_5 = -0.6$, and derive exact solution of rogue wave for Eq. (1) as follows[28]

$$Q(t,x) = \frac{e^{it}(-0.5x^2 - 0.5t^2 + ti + 0.6xt - 0.18t^2 + 0.375)}{0.5t^2 + 0.5x^2 - 0.6xt + 0.18t^2 + 0.125}, \quad x \in [-2,2],\ t \in [-1.5, 1.5], \quad (16)$$

and thus

$$Q(-1.5, x) = \frac{e^{-1.5i}(-0.5x^2 - 1.155 - 1.5i - 0.9x)}{1.655 + 0.5x^2 - 0.9x}, \quad x \in [-2, 2], \quad (17)$$

with the Dirichlet-Neumann periodic boundary condition

$$Q(t, -2) = Q(t, 2), \quad t \in [-1.5, 1.5] \quad (18)$$

The sampling points is obtained by means of pseudo-spectral method with space-time region $(x,t) \in [-2,2] \times [-1.5, 1.5]$, and the rogue wave solution is discretized into $[513 \times 401]$ data points. Similar to the procedure of one-soliton, after 4000 iterations of learning, the network achieved a relative error $L_2$ of $2.555952 \times 10^{-2}$ in about 47319.1462 seconds.

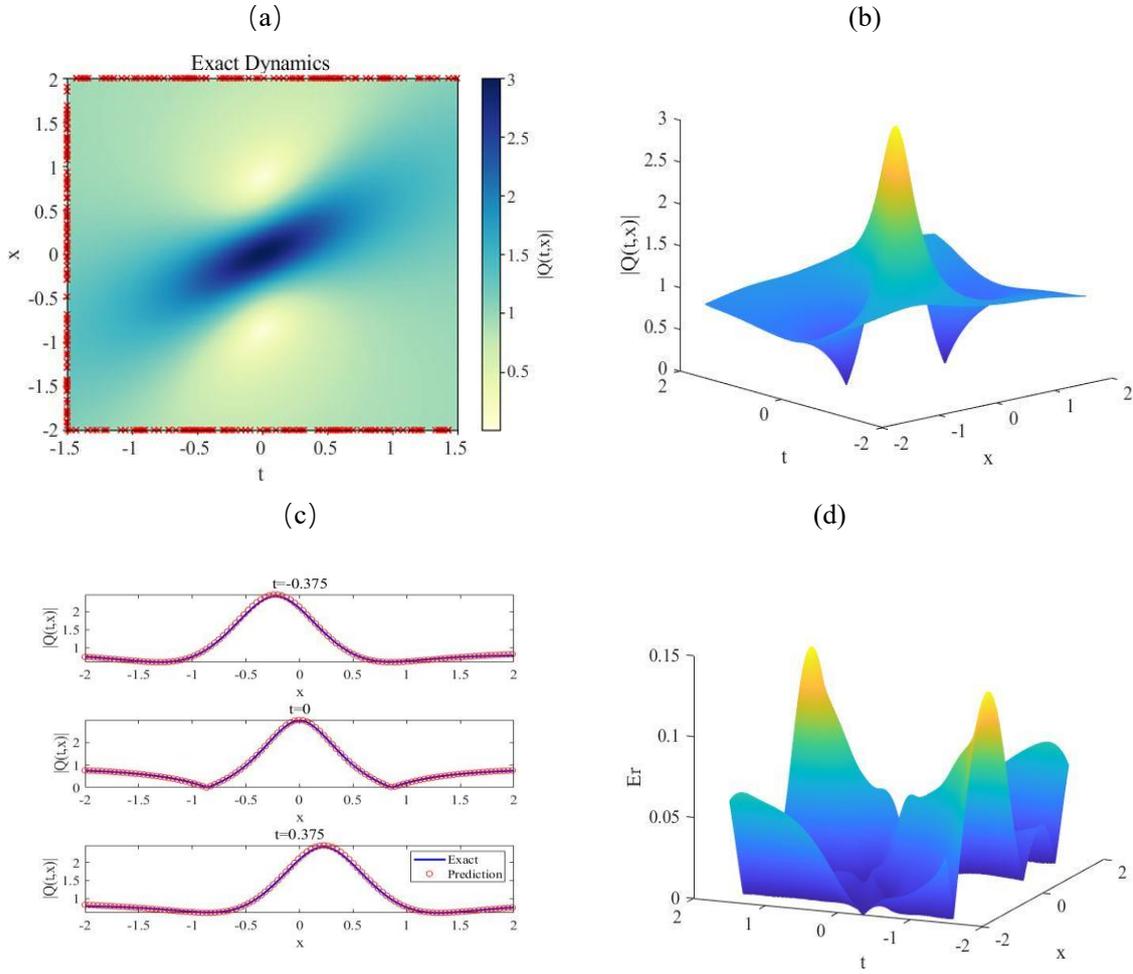

**Figure 3.** Rogue wave solution $Q(t,x)$: (a) Exact solution $|Q(t,x)|$ with the boundary and initial sampling points; (b) Reconstruct spatiotemporal dynamics of rogue wave solution; (c) Comparison between the exact and predicted solutions at three distances, with the red hollow dots as the predicted values and the blue solid lines as the exact values; (d) The three-dimensional stereogram of error.

Fig. 3 exhibits space-time dynamics of reconstructed rogue wave. In Fig. 3(a), the sampling point configuration of the initial and boundary is clearly analyzed. Fig. 3(b) reconstructs the space-time dynamics of the rogue wave for the HNLSE. In Fig. 3(c), we compare the predicted solution with exact solution at three different distances $t = -0.375, 0, 0.375$. Fig. 3(d) shows the Error $Er = |\hat{Q}(t,x) - Q(t,x)|$ between the predicted and exact solutions with an error below $2 \times 10^{-1}$, and thus these predictions are relatively accurate.

### 3.4 W-soliton solution

We take the parameters $\lambda_1 = 0$, $\lambda_2 = 0.1$, $\lambda_3 = 0$, $\lambda_4 = 0.01$ and $\lambda_5 = -0.015$ in Eq. (1), and exact solution of W-soliton is as follows [31]

$$Q(t,x) = i - i(1+\sqrt{2})\operatorname{sech}(x-t), \quad x \in [-10,10], \quad t \in [0,5], \tag{19}$$

with the initial conditions

$$Q(0,x) = i(1 - (1+\sqrt{2})\operatorname{sech}(x)), \quad x \in [-10,10], \tag{20}$$

and the periodic condition

$$Q(t,-10) = Q(t,10), \quad t \in [0,5]. \tag{21}$$

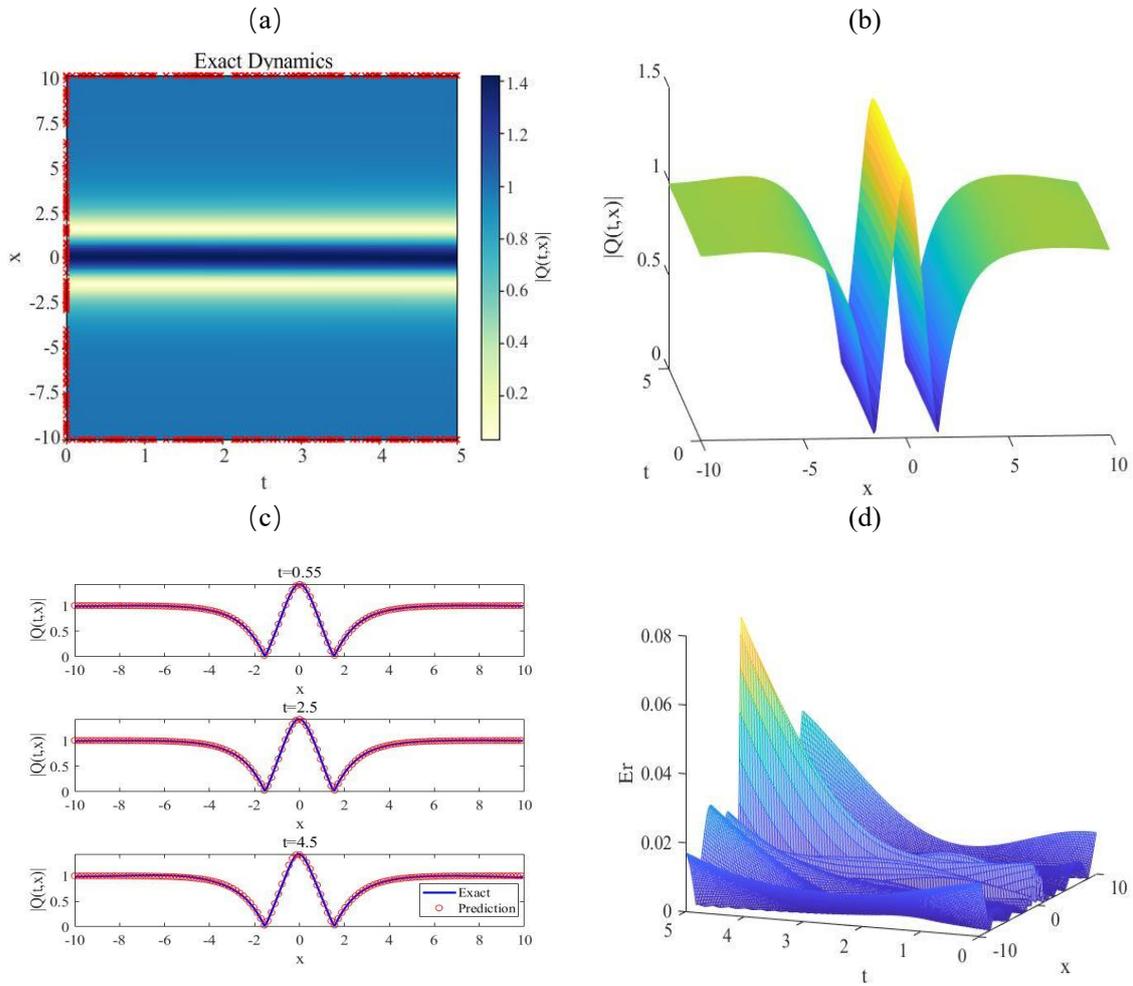

**Figure 4.** W-soliton solution $Q(t,x)$: (a) Exact solution $|Q(t,x)|$ with the boundary and initial sampling points; (b) Reconstructed W-soliton space-time dynamics; (c) Comparison between the exact and predicted solutions at three distances, with the red hollow dots as the predicted values and the blue solid lines as the exact values; (d) The three-dimensional stereogram of error.

The sampling points set is obtained by means of pseudo-spectral method with space-time region $(x,t) \in [-10,10] \times [0,5]$, and the W-soliton solution is discretized into $[256 \times 201]$ data points. Similar to the procedure of one-soliton, the MSE loss function given by Eq. (5) is learned by using a PINN consisting of 7 hidden layers with 40 neurons in each layer. After 4000 iterations of learning, the network achieved a relative error $L_2$ of $9.128569 \times 10^{-3}$ in about 3826.5934 seconds.

Fig. 4 displays space-time dynamics of reconstructed W-soliton. In Fig. 4(a), the sampling point configuration of the initial and boundary is clearly analyzed. Fig. 4(b) reconstructs the space-time dynamics of the W-soliton for the HNLSE. In Fig. 4(c), we compare the predicted solution with exact solution at three different distances $t = 0.55, 2.5, 4.5$. Fig. 4 (d) shows that the error between the predicted and exact solutions is very small, and the predicted result is accurate in the whole time and space domains.

### 3.5 M-soliton solution

We take the parameters $\lambda_1 = 0$, $\lambda_2 = 2$, $\lambda_3 = 0$, $\lambda_4 = 0.0247$ and $\lambda_5 = -0.03705$ in Eq. (1), and the exact solution of M-soliton is as follows [32]

$$Q(t,x) = \mathrm{sech}(0.5(x-t)) \tanh(x-t) e^{i(t-80.97x)}, \quad x \in [-10,10], \quad t \in [0,5]. \tag{22}$$

To use the PINN deep learning, we consider the initial condition

$$Q(0,x) = \mathrm{sech}(0.5x) \tanh x \, e^{-80.97xi}, \quad x \in [-10,10], \tag{23}$$

and the Dirichlet-Neumann periodic boundary condition

$$Q(t,-10) = Q(t,10), \quad t \in [0,5]. \tag{24}$$

Similar to the procedure of W-soliton, after 4000 iterations of learning, the network achieved a relative error $L_2$ of $6.015803 \times 10^{-3}$ in about 7924.9628 seconds. Fig. 5 displays space-time dynamics of reconstructed M-soliton. In Fig. 5(a), the sampling point configuration of the initial and boundary is clearly analyzed. Fig. 5(b) reconstructs the space-time dynamics of the M-soliton for the HNLSE. In Fig. 5(c), we compare the predicted solution with exact solution at three different distances $t = 0.55, 2.5, 4.5$. Fig. 5 (d) represents the density map of the error between the predicted and exact solution. From the density map, it can be seen that the prediction results of the optimized PINN are very accurate globally.

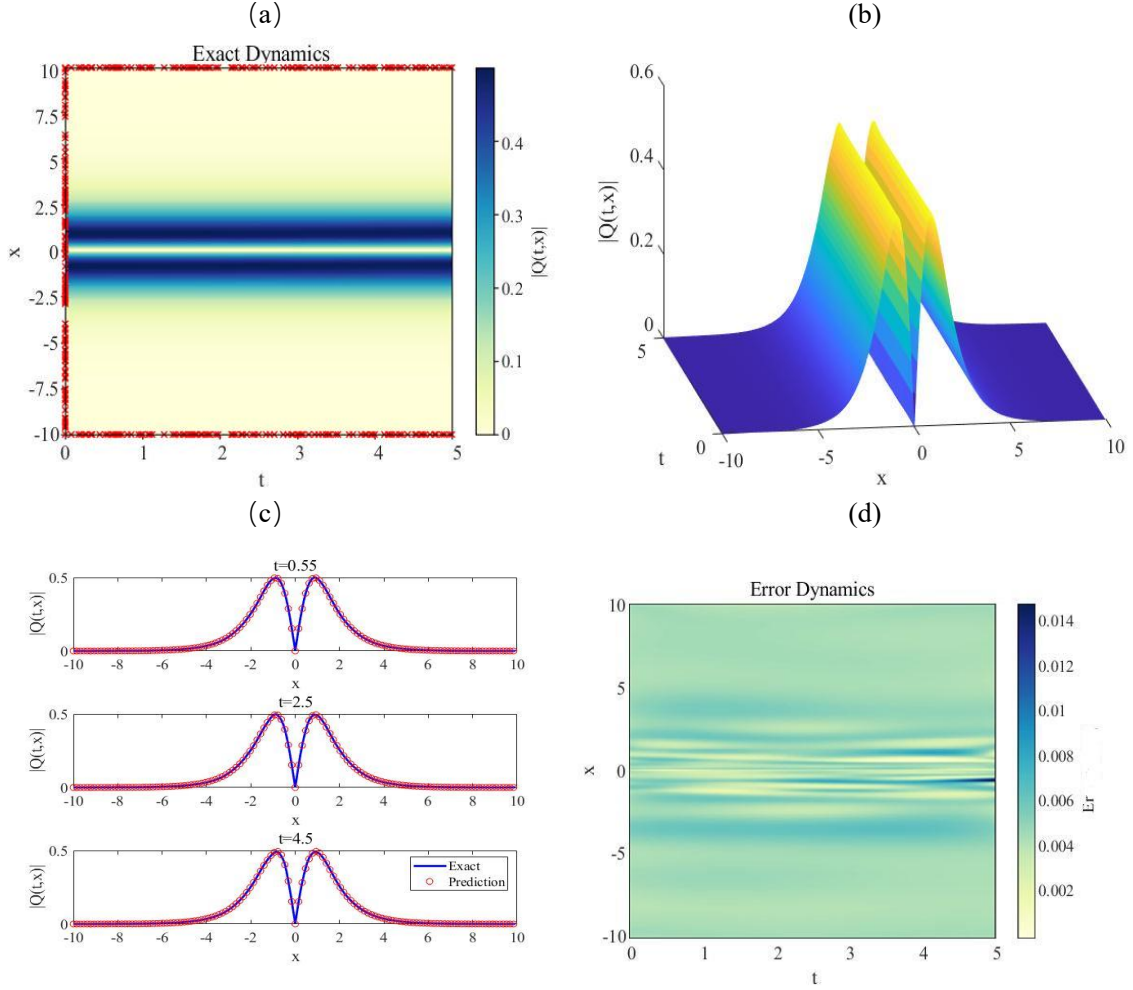

**Figure 5.** M-soliton solution $Q(t,x)$: (a) Exact solution $|Q(t,x)|$ with the boundary and initial sampling points; (b) Reconstructed M-soliton space-time dynamics; (c) Comparison between the exact and predicted solutions at three distances, with the red hollow dots as the predicted values and the blue solid lines as the exact values; (d) The error density plot.

## 4. Data-driven parameter discovery of HNLSE

In this section, we considered the data-driven parameters discovery of the HNLSE using the PINN. The HNLSE is as follows

$$iQ_t + \lambda_1 Q_{xx} + \lambda_2 |Q|^2 Q + i(\lambda_3 Q_{xxx} + \lambda_4 |Q|^2 Q_x) = 0, \tag{25}$$

where the pulse envelope $Q = u(x,t) + iv(x,t)$ with the real and imaginary parts $u$, $v$, coefficients $\lambda_1, \lambda_2, \lambda_3, \lambda_4$ are unknown parameters that need to be determined by the PINN training.

The network $f(x,t)$ is defined as

$$f := iQ_t + \lambda_1 Q_{xx} + \lambda_2 |Q|^2 Q + i(\lambda_3 Q_{xxx} + \lambda_4 |Q|^2 Q_x), \tag{26}$$

and the $f(x,t)$ has both real and imaginary parts as $f(x,t) = f_u(x,t) + if_v(x,t)$, thus

$$\begin{aligned} f_u &:= u_t + \lambda_1 v_{xx} + \lambda_2 (u^2 + v^2)v + \lambda_3 u_{xxx} + \lambda_4 u_x (u^2 + v^2) \\ f_v &:= v_t - \lambda_1 u_{xx} - \lambda_2 (u^2 + v^2)u + \lambda_3 v_{xxx} + \lambda_4 v_x (u^2 + v^2), \end{aligned} \tag{27}$$

to build a multi-output neural network with the output being the complex value of network $f(x,t)$. These unknown parameters in PINN can be learned by minimizing the MSE loss function

$$LOSS = \frac{1}{N_f}\sum_{j=1}^{N_f}(|u(x^j,t^j)-u^j|^2 + |v(x^j,t^j)-v^j|^2 + |f_u(x^j,t^j)|^2 + |f_v(x^j,t^j)|^2). \tag{28}$$

By using the pseudo-spectral method, exact rogue wave solution for the HNLSE (25) with parameters $\lambda_1 = 0.5, \lambda_2 =1, \lambda_3 = 0.1, \lambda_4 = 0.6$ is considered to make sampling pointsets with space-time region $(x,t) \in [-8,8] \times [-2,2]$, and exact rogue wave solution is discretized into $[256 \times 201]$ data points.

**Table 1** The correct HNLSE and the identified equation obtained by learning the unknown parameter $\lambda_1$ and $\lambda_2$.

| Correct HNLSE | $iQ_t + 0.5Q_{xx} + |Q|^2 Q + i(0.1Q_{xxx} + 0.6|Q|^2 Q_x) = 0$ | Error |
|---|---|---|
| Identified HNLSE (clean data) | $iQ_t + 0.50004Q_{xx} + 1.0003|Q|^2 Q + i(0.1Q_{xxx} + 0.6|Q|^2 Q_x) = 0$ | $\lambda_1 = 0.00739\%, \lambda_2 = 0.00254\%$ |
| Identified HNLSE (1% noise) | $iQ_t + 0.49982Q_{xx} + 0.99982|Q|^2 Q + i(0.1Q_{xxx} + 0.6|Q|^2 Q_x) = 0$ | $\lambda_1 = 0.03508\%, \lambda_2 = 0.01777\%$ |

**Table 2** The correct HNLSE and the identified equation obtained by learning the unknown parameter $\lambda_1, \lambda_2$ and $\lambda_3$.

| Correct HNLSE | $iQ_t + 0.5Q_{xx} + |Q|^2 Q + i(0.1Q_{xxx} + 0.6|Q|^2 Q_x) = 0$ | Error |
|---|---|---|
| Identified HNLSE (clean data) | $iQ_t + 0.50006Q_{xx} + 1.0004|Q|^2 Q + i(0.10003Q_{xxx} + 0.6|Q|^2 Q_x) = 0$ | $\lambda_1 = 0.01121\%, \lambda_2 = 0.00397\%$ $\lambda_3 = 0.02848\%$ |
| Identified HNLSE (1% noise) | $iQ_t + 0.50001Q_{xx} + 0.99997|Q|^2 Q + i(0.09995Q_{xxx} + 0.6|Q|^2 Q_x) = 0$ | $\lambda_1 = 0.00177\%, \lambda_2 = 0.00311\%$ $\lambda_3 = 0.05355\%$ |

**Table 3** The correct HNLSE and the identified equation obtained by learning the unknown parameter $\lambda_1, \lambda_2, \lambda_3$ and $\lambda_4$.

| Correct HNLSE | $iQ_t + 0.5Q_{xx} + |Q|^2 Q + i(0.1Q_{xxx} + 0.6|Q|^2 Q_x) = 0$ | Error |
|---|---|---|
| Identified HNLSE (clean data) | $iQ_t + 0.50008Q_{xx} + 0.99997|Q|^2 Q + i(0.10001Q_{xxx} + 0.59992|Q|^2 Q_x) = 0$ | $\lambda_1 = 0.01578\%, \lambda_2 = 0.00308\%$ $\lambda_3 = 0.01290\%, \lambda_4 = 0.01366\%$ |
| Identified HNLSE (1% noise) | $iQ_t + 0.50022Q_{xx} + 1.00009|Q|^2 Q + i(0.09971Q_{xxx} + 0.59848|Q|^2 Q_x) = 0$ | $\lambda_1 = 0.04470\%, \lambda_2 = 0.00870\%$ $\lambda_3 = 0.28720\%, \lambda_4 = 0.25308\%$ |

In order to learn these unknown parameters $\lambda_1, \lambda_2, \lambda_3$ and $\lambda_4$ via the PINN, the number of samples is randomly selected from the sampling points $N_f = 5000$, and 6-layers of deep neural network and network structure of 50 neurons in each layer is chosen to learn the PINN $f(x,t)$. Tables 1-3 illustrate the training results of unknown parameters in different situations, and show the training error is given at the last column. In Table 1, parameters $\lambda_3$ and $\lambda_4$ are fixed as 0.1 and 0.6 in the neural network, and parameter $\lambda_4$ is fixed as 0.6 in the neural network in Table 2. We find that even if the sampling points is destroyed by noise, the PINN can correctly identify unknown parameters and give very high accuracy. The network can recognize HNLSE with significant accuracy, even if the sampling points is corrupted by 1% irrelevant noise. Meanwhile, with the gradual increase of training parameters, the error will gradually add, but these results are still robust.

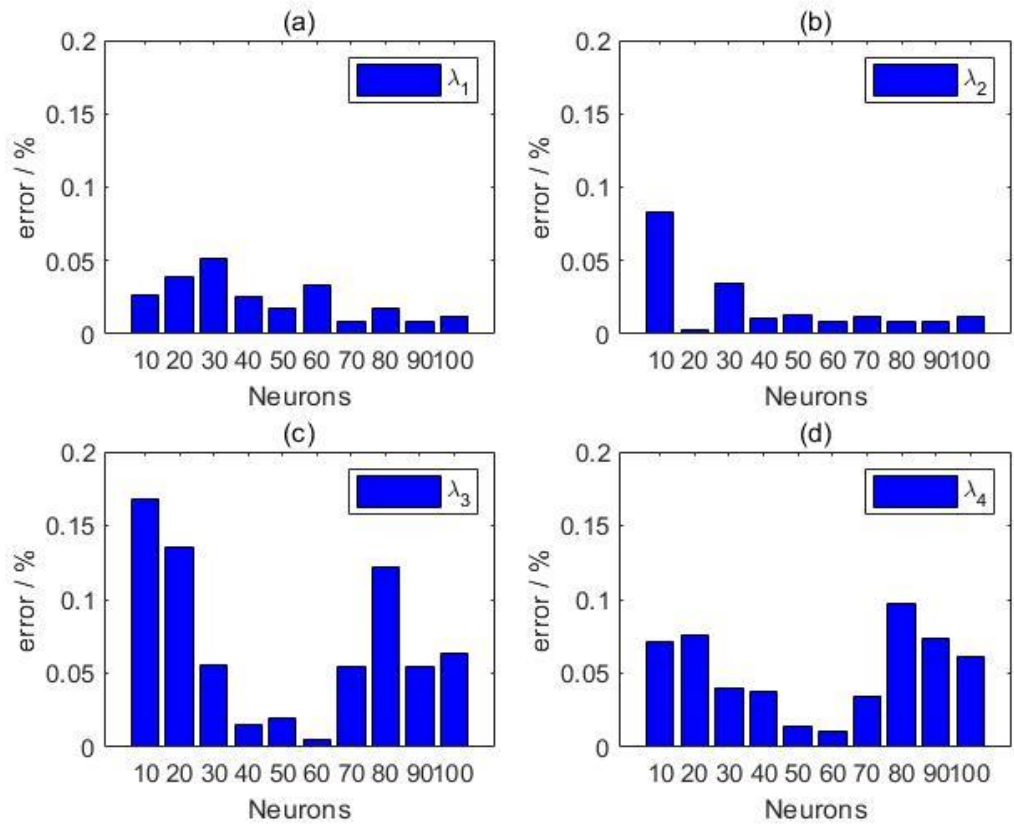

**Figure 6.** Training errors of unknown parameters ($\lambda_1, \lambda_2, \lambda_3, \lambda_4$) with different number of neurons in each layer. The neural network structure is chosen as 8 layers, and the sampling points $N_f = 5000$.

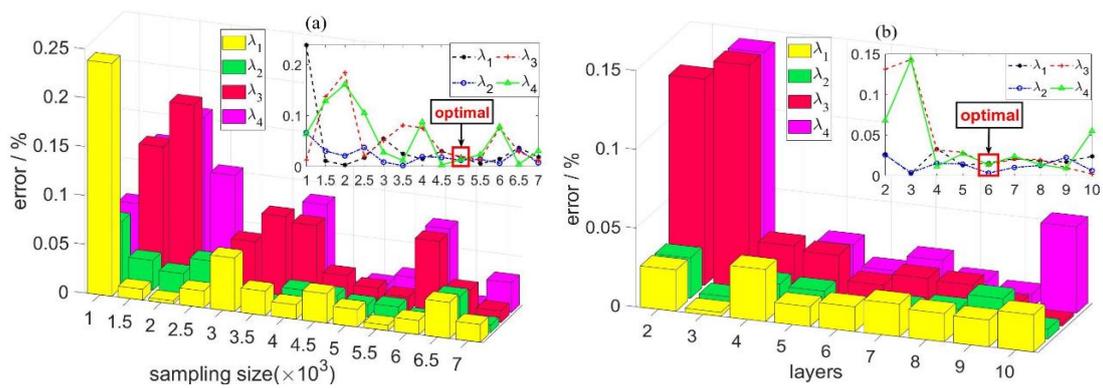

**Figure 7.** Training errors of unknown parameters ($\lambda_1, \lambda_2, \lambda_3, \lambda_4$) in different sampling sizes and hidden layers with (a) 8 hidden layers with 50 neurons in each layer, and (b) 50 neurons in each layer. Here the sampling points $N_f = 5000$.

Under the condition of a small amount of sampling points, our main purpose is to predict the important physical parameters of the model and make our error further reduced. However, our optimization for HNLSE may be effective for this model, but not for other models, so we need to find a more general rule. This will be our next research direction.

In order to further analyze how to reduce the training error, we conducted a systematic study on the total number of sampling points, the number of neurons and the number of hidden layers of neural networks. We use the control variable method to find an optimal structure for the neural network. These results are shown in Fig. 6 and 7. Even if the number of neurons changes, the training results are still stable. By observing the training results of four unknown parameters, we find 50 neurons per layer is a suitable choice considering the operation speed of neural network. In Fig 7, by changing the randomly sampled sampling points and the number of hidden layers, we find that the sampling points as $N_f = 5000$ and the hiding layer as 6 will be a better choice. Compared with previous literatures on data-driven parameter discovery [18,24], our error is greatly reduced. Similar training results can be obtained by using bright solitons and other exact solutions as sampling points.

Of course, our neural network structure is only a local optimal choice, and we are not sure whether there will be better results if we continue to increase these three variables, that is, the total number of sampling points, the number of neurons and the number of hidden layers of neural networks. We hope to find a general rule in the next research.

5. **Conclusion**

In conclusion, we study the one-soliton solution, two-soliton solution, rogue wave solution, W-soliton solution and M-soliton solution of HNLSE via the multi-level PINN deep learning method under different initial and periodic boundary conditions. In particular, when we use the network to train W-soliton and M-soliton, we find that the relative error $L_2$ of the predicted results reaches $1\times10^{-3}$, and the training speed is fast. The results show that the training effect for simple soliton is better. In Figs.4(d) and 5(d), as the learning distance increases, the error gradually increases. Since the sampling position is on the initial boundary, as the learning distance increases, the error between the predicted result and the actual situation will gradually increase.

In addition, based on rogue wave solution, the data-driven parameter discovery of HNLSE is studied. We add the noise to judge the stability of the neural network. The prediction error will increase as the number of unknown parameters adds. These changes of error are within the controllable range, which is enough to prove the excellent performance of PINN. We have optimized the structure of the neural network mainly from three aspects including the number of layers of the neural network, the number of neurons, and the sampling points. Compared with previous studies [24], our error is greatly reduced.

**Acknowledgements**

This work is supported by the Zhejiang Provincial Natural Science Foundation of China (Grant No. LR20A050001) and the National Natural Science Foundation of China (Grant Nos. 12075210 and 11874324).

**Conflict of interest**

The authors have declared that no conflict of interest exists.

**Ethical Standards**

This Research does not involve Human Participants and/or Animals.